\documentclass[english,manuscript,superscriptaddress]{revtex4-1}
\usepackage[T1]{fontenc}
\usepackage[latin9]{inputenc}
\usepackage{babel}

\usepackage{graphicx}
\usepackage[unicode=true, pdfusetitle,
 bookmarks=true,bookmarksnumbered=false,bookmarksopen=false,
 breaklinks=false,pdfborder={0 0 1},backref=false,colorlinks=false]
 {hyperref}

\makeatletter

\providecommand{\tabularnewline}{\\}
\newcommand{\lyxdot}{.}

 \@ifundefined{textcolor}{}
 {%
   \definecolor{BLACK}{gray}{0}
   \definecolor{WHITE}{gray}{1}
   \definecolor{RED}{rgb}{1,0,0}
   \definecolor{GREEN}{rgb}{0,1,0}
   \definecolor{BLUE}{rgb}{0,0,1}
   \definecolor{CYAN}{cmyk}{1,0,0,0}
   \definecolor{MAGENTA}{cmyk}{0,1,0,0}
   \definecolor{YELLOW}{cmyk}{0,0,1,0}
 }

\makeatother

\makeatother

\begin{document}

\title{A search for b(c) quark pdf uncertainties at TeV scale $ep$ collider}

\author{H. Aksakal}

\email{aksakal@cern.ch}

\affiliation{Physics Department, Art and Science Faculty Nigde University, Nigde,
Turkey}

\author{S. O. Kara}

\email{seyitokankara@gmail.com}

\affiliation{Bor Vocational School, Nigde University, Nigde,Turkey}
\begin{abstract}
We discuss $c\bar{c}$ and $b\bar{b}$ pair productions at $ep$ collider
for studying extremely small $x(g)$ region. It has been shown that
Large Hadron electron Collider (LHeC) has a reach of about $x(g)>10^{-6}$.
The aim of this work is to show that the PDF uncertainties in the
heavy flavour production. Maximum difference of cross section between
PDFs 60\% has been found in the process of {\normalsize $ep\rightarrow eq\bar{q}X$.} 
\end{abstract}
\maketitle

\section*{1. Introduction}

The Large Hadron Collider (LHC) will provide unique physics opportunities
for SM and BSM physics. In hadron colliders, one of the sources of
the systematic errors on the measured quantities is the uncertainty
in the parton distribution functions (PDFs). The current proton PDF
knowledge mostly originates from the deep inelastic scattering (DIS)
measurements at the first $ep$ collider, HERA. It has probed small
$x(g)$ region and had a reach of about $x(g)>10^{-4}$. Probing even
smaller $x(g)$ region \cite{key-1} could be realized with the Large
Hadron Electron Collider (LHeC) project \cite{key-2}. In LHeC, electron
beams of $60$ and $140$ GeV energy accelerated either by a linac
(linac-ring version) or by the ring installed in LHC tunnel (ring-ring
version) are collided with $7$ TeV LHC protons (or ions).

The photon-gluon fusion in LHeC is to produce heavy flavour quark
pairs as it can be seen in Fig.1. In this vertex the gluon flavour
PDF uncertainty has to be considered\cite{key-2}. The gluon PDF is
responsible to determine small $x(g)$ region for low momentum fraction.
In the conventional QCD framework, the PDFs for charm ($c$) and bottom
($b$) quarks are determined by fitting to the hadronic data \cite{key-3}.
In this study the maximum values of differential cross section and
corresponding $x(g)$ values have been investigated for several different
PDFs in CompHEP \cite{key-4} software package. Obtained results clearly
show the difference betwen PDFs. The investigation of small $x(g)$
region using the processes $ep\rightarrow eb\bar{b}X$ and $ep\rightarrow ec\bar{c}X$
is presented in section 2. In section 3 the accelerator properties
of LHeC have been summarized. Finally, the generator level results
of PDF distributions with $b\bar{b}$ and $c\bar{c}$ pair production
obtained using CompHEP are analyzed in section 4.

\section*{2. The Physics Case}

In the $ep$ option of the LHeC one can consider two cases: firstly
e-beam energy with $60$ GeV (LHeC Type-1) and second $140$ GeV (LHeC
Type-2). On the other hand, in another version of the $ep$ collider,
the beam energies can be extended to $250$ GeV (and/or $500$ GeV).
The processes $ep\rightarrow eb\bar{b}X$ and $ep\rightarrow ec\bar{c}X$
have been used in the PDF uncertainty studies. The subprocesses $eg\rightarrow eb\bar{b}$
and $eg\rightarrow ec\bar{c}$ have been used to measure of the $x(g)$
where the gluon ($g$) is from the LHC protons, electrons are from
an electron linac. The $b$ quark final states are easier to identify
due to $b$-tagging possibility using currently available technologies:
for example, ATLAS silicon detectors have $70\%$ b-tagging efficiency\cite{key-5}.

In Table 1 we introduce the cross sections of heavy quark pair production
for LHeC and $ep$ colliders. CTEQ6L1\cite{key-6,key-7} PDF distribution
in CompHEP simulation program has been chosen for all calculations
in Table 1.

\begin{table}
\begin{tabular}{|c|c|c|}
\hline 
LHeC and ep collider  & $b\bar{b}$ ($pb$)  & $c\bar{c}$ ($pb$)\tabularnewline
\hline
\hline 
LHeC Type-1($E_{e}=60$GeV)  & $4.19\times10^{3}$  & $2.36\times10^{5}$\tabularnewline
\hline 
LHeC Type-2($E_{e}=140$GeV)  & $6.95\times10^{3}$  & $3.67\times10^{5}$\tabularnewline
\hline 
$ep$ collider-1($E_{e}=250$GeV)  & $9.65\times10^{3}$  & $4.91\times10^{5}$\tabularnewline
\hline 
$ep$ collider-2($E_{e}=500$GeV)  & $1.4\times10^{4}$  & $6.89\times10^{5}$\tabularnewline
\hline
\end{tabular}\caption{Heavy quark pair production cross sections for LHeC and $ep$ collider}

\end{table}

In Table 2, maximum values of differential cross sections and corresponding
$x(g)$ values are given for different PDF distributions at LHeC and
$ep$ collider. For example the differential cross sections for LHeC
Type-1 and LHeC Type-2 achieve maximum values at CTEQ5L, while maximum
values for $ep$ collider-1 and $ep$ collider-2 are CTEQ4L and CTEQ5L
for $b\bar{b}$, CTEQ6L1 for $c\bar{c}$, respectively.

\begin{table}
\begin{tabular}{|c|c|c|c|c|}
\hline 
LHeC and ep collider  & \multicolumn{2}{c|}{$b\bar{b}$} & \multicolumn{2}{c|}{$c\bar{c}$ }\tabularnewline
\hline 
 & $d\sigma/dx$  & $x$  & $d\sigma/dx$  & $x$\tabularnewline
\hline 
LHeC Type-1  & $0.49nb$  & $2.15\times10^{-4}$  & $0.13\mu b$  & $3.53\times10^{-5}$\tabularnewline
\hline 
LHeC Type-2  & $1.77nb$  & $9.17\times10^{-5}$  & $0.45\mu b$  & $1.7\times10^{-5}$\tabularnewline
\hline 
ep collider-1  & $4.19nb$  & $4.88\times10^{-5}$  & $1.06\mu b$  & $9.95\times10^{-6}$\tabularnewline
\hline 
ep collider-2  & $10.62nb$  & $2.75\times10^{-5}$  & $2.6\mu b$  & $5.15\times10^{-6}$\tabularnewline
\hline
\end{tabular}

\caption{Maximum values of $d\sigma/dx(g)$ and corresponding $x(g)$ values
for $b\bar{b}$ and $c\bar{c}$ final states at LHeC and $ep$ collider.}

\end{table}

\begin{figure}
\includegraphics[scale=0.4]{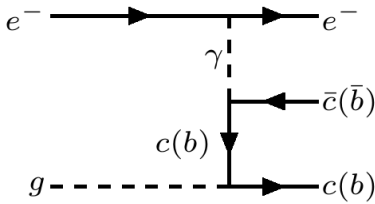}\caption{The tree level Feynman diagram for heavy quark pair production}

\end{figure}

\section*{3. The Accelerator Properties}

The LHeC can be obtained by using a linac to accelerate $e$$(e^{+}$or
$\gamma)$-beam to 60-140 GeV and colliding with 7 TeV LHC proton
beam most realistically from the upgraded LHC. Such a collider has
been proposed previously under the name of {}``QCD Explorer\textquotedblright{}
project \cite{key-10,key-11}. The linac would be based on superconducting
RF technology\cite{key-2,key-12}. The Table III contains the Linac
and LHC parameters. To provide the $e$-beam, so far two accelerator
options have been considered. A {}``ring\textquotedblright{} option
that can be achieved by the installation of an additional $e^{-}$(or
$e^{+}$) ring inside the LHC tunnel\cite{key-13}. This option has
the benefit of having the circulating beams similar to well known
LEP operations however the maximum energy that could be provided by
the $ee^{+}$ beams is rather limited. There are two version of linac
option under consideration: multipass energy recovery linac (ERL)
and the second a pulsed linac. First one has $10^{33}$ luminosity
can not reach more than $60$ GeV beam energy because of rigorous
synchrotron radiation. In the second option, the e-beam energy can
be obtained by the utilization of linac\cite{key-2}. The second option
have two possibilities which are: an ILC like SC linac and a NC linac.
Efficient positron production for $e^{+}p$ collision at LHeC is explained
in Ref.\cite{key-14}. $\gamma-p$ collision at LHeC is also possible
where $\gamma$ comes from Compton backscattering of laser photons
off $e$-beam and it is an advantage for Linac-ring option but not
possible for the ring-ring one\cite{key-15,key-16}. An additional
benefit of a linac is the possibility of high electron (thus photon)
polarization. The luminosity of the collider has been calculated by
CAIN code \cite{key-17} which was originally written for $e^{-}(e^{+}$or
$\gamma)$$-$$e^{-}(e^{+}$or $\gamma)$ collision and can be easily
adopted to $ep$ collision. In the case of LHC upgraded option of
Large Pwinski Angle (LPA) \cite{key-12} is used then number of proton
per bunch can be inreased by 2.5 times thus the luminosity is also
inreased by a factor of 2.5.

\begin{table}
\begin{tabular}{|c|cccc|c|}
\hline 
parameter  & \multicolumn{4}{c|}{Linac (ERL /pulsed(SC))} & LHC\tabularnewline
\hline
\hline 
RF Frequency $F_{RF}$($GHz$)  & \multicolumn{4}{c|}{$0.721/1.3$ } & $0.4$\tabularnewline
\hline 
Beam energy $E_{b}$ ($GeV$)  & \multicolumn{4}{c|}{$60/140-250-500$ } & $7000$\tabularnewline
\hline 
Bunch length $\sigma_{z}$ ($mm$)  & \multicolumn{4}{c|}{$0.3$ } & $75.5$\tabularnewline
\hline 
Bunch size at IP $\sigma_{x}/\sigma_{y}$ ($\mu m$)  & \multicolumn{4}{c|}{$7$} & $7$\tabularnewline
\hline 
Tr. nor. emittance $\gamma\epsilon_{x}/\gamma\epsilon_{y}$ ($\mu m$)  & \multicolumn{4}{c|}{$50$} & $3.75$\tabularnewline
\hline 
Bunch spacing ($ns$)  & \multicolumn{4}{c|}{$25(\mbox{or}50)/25(\mbox{or}50)$} & $25$ (or$50$)\tabularnewline
\hline 
Rep. Freq. ($Hz$) & \multicolumn{4}{c|}{$40(\mbox{or }20)\,10^{6}/10\,(\mbox{or }5)$} & NA\tabularnewline
\hline 
\# of bunch & \multicolumn{4}{c|}{$CW/1\:10^{5}$} & $2808$ (or$1404$)\tabularnewline
\hline 
Bunch length ($mm$) & \multicolumn{4}{c|}{$0.3$} & $75.5$\tabularnewline
\hline 
Bunch population  & \multicolumn{4}{c|}{$1\mbox{\ensuremath{(}or}$$2$)$10^{9}$ /$2$$10^{9}$} & $1.7$$10^{11}$\tabularnewline
\hline 
Average beam current $I$ ($mA$)  & \multicolumn{4}{c|}{$6.4$$/3.2$} & $>430$\tabularnewline
\hline 
$\sqrt{s}$ $TeV$  & \multicolumn{1}{c}{$1.3$ } & $1.98$  & $2.64$  & \multicolumn{2}{c|}{$3.74$}\tabularnewline
\hline 
Luminosity $10^{32}$($cm$$^{-2}s^{-1}$)  & 9.7 & 0.68 & 0.78 & \multicolumn{2}{c|}{0.87}\tabularnewline
\hline
\end{tabular}\caption{Linac and LHC Parameters}

\end{table}

\section*{4. Physics Search Potential and PDF Uncertainty}

The precise measurement of PDFs play crucial role in the framework
of QCD studies. The different differential cross sections obtained
from different PDFs originate in PDF uncertainties. Used PDFs can
be divided in three groups which are leading order-LO (cteq5l, cteq4l,
cteq6l), next-to-leading order-NLO (cteq5m1, cteq6d), and $\overline{MS}$
(cteq4m, cteq6m, cteq6l1). Perturbative correction in DIS (pQCD) includes
LO and NLO contribution. $\overline{MS}$ scheme distributions defined
by matrix elements are not simple one loop in perturbation theory
and for most application $\overline{MS}$ parton distribution is mostly
used \cite{key-18,key-19,key-20}. As seen from Fig.2, the difference
is quite evident for large values of $\sqrt{s}$, whereas it is closely
for the center of mass energy less than 1 TeV. Furthermore this figure
shows two groups of curves for the larger $\sqrt{s}$ values: the
differential cross sections of first group are around at $25$ $nb$
and second at $15$ $nb$.

In Fig. 3 differential cross section versus the $x(g)$ reach at LHeC
Type-1 is plotted for different PDFs. It is seen that the $d\sigma/dx(g)$
of $c\bar{c}$ pair production is larger than of $b\bar{b}$ pair
production. In both final states, CTEQ6L1, CTEQ5L and CTEQ4L adopt
a similar manner for large values of $d\sigma/dx(g)$ and the others
without CTEQ6L for small values. $d\sigma/dx(g)$ of CTEQ6L achieves
maximum value $0.4$ $nb$ for $b\bar{b}$ and $0.1$ $\mu b$ for
$c\bar{c}$ pair productions.

Similar distributions for $b\bar{b}$ and $c\bar{c}$ pair productions
at LHeC Type-2 are shown in Fig. 4. For example, differential cross
section of $b\bar{b}$ pair production for CTEQ5L achieves maximum
value $1.77$ $nb$ at $x(g)=9.17\times10^{-5}$ whereas that of $c\bar{c}$
pair production for CTEQ5L is $0.45$ $\mu b$ at $x(g)=1.7\times10^{-5}$.

In order to show the distributions of PDFs for $ep$ collider-1 we
present the Fig. 5, where the shift in the $x(g)$ values of some
PDF peaks from the others is clearly seen for $c\bar{c}$ final state.
The same shift for $c\bar{c}$ pair production at $ep$ collider-2
is more explicitly seen in Fig. 6. This shift comes from gluon uncertainties
which is most uncertain of PDFs and it is increases with $\sqrt{s}$
, can be used to choose more appropriate PDF when the experimental
data became availible.

\section*{5. Conclusions}

By investigating the processes $ep\rightarrow eb\bar{b}X$ and $ep\rightarrow ec\bar{c}X$,
we have shown that PDFs for LHeC and an $ep$ collider adopt different
manner. Obtained peak difference of differential cross sections in
between these PDFs is about $60\%$ and it is arise from $\mbox{LO}$,
NLO and $\overline{MS}$ contribution. The difference can be reduced
by taken into account unimplemented corrections because total cross
section of DIS is not simple finite function of $\alpha_{s}$, those
unimplemented corrections will be searched as a next publication.
As a result of calculations done, we could say that the differential
cross sections obtained from different PDFs are closely for the center
of mass energy less than $1$ TeV, whereas they are quite evidently
for large values of $\sqrt{s}$. The differential cross sections of
each final states at LHeC and $ep$ collider achieve maximum values
at the different PDF. Biggest uncertainity of PDFs is found in$c\bar{c}$
final state at $\sqrt{s}=3.74$ TeV.
\begin{acknowledgments}
The authors would like to thank S. Sultansoy and G. Unel for useful
discussion.\end{acknowledgments}

\begin{figure}
\includegraphics[scale=0.7]{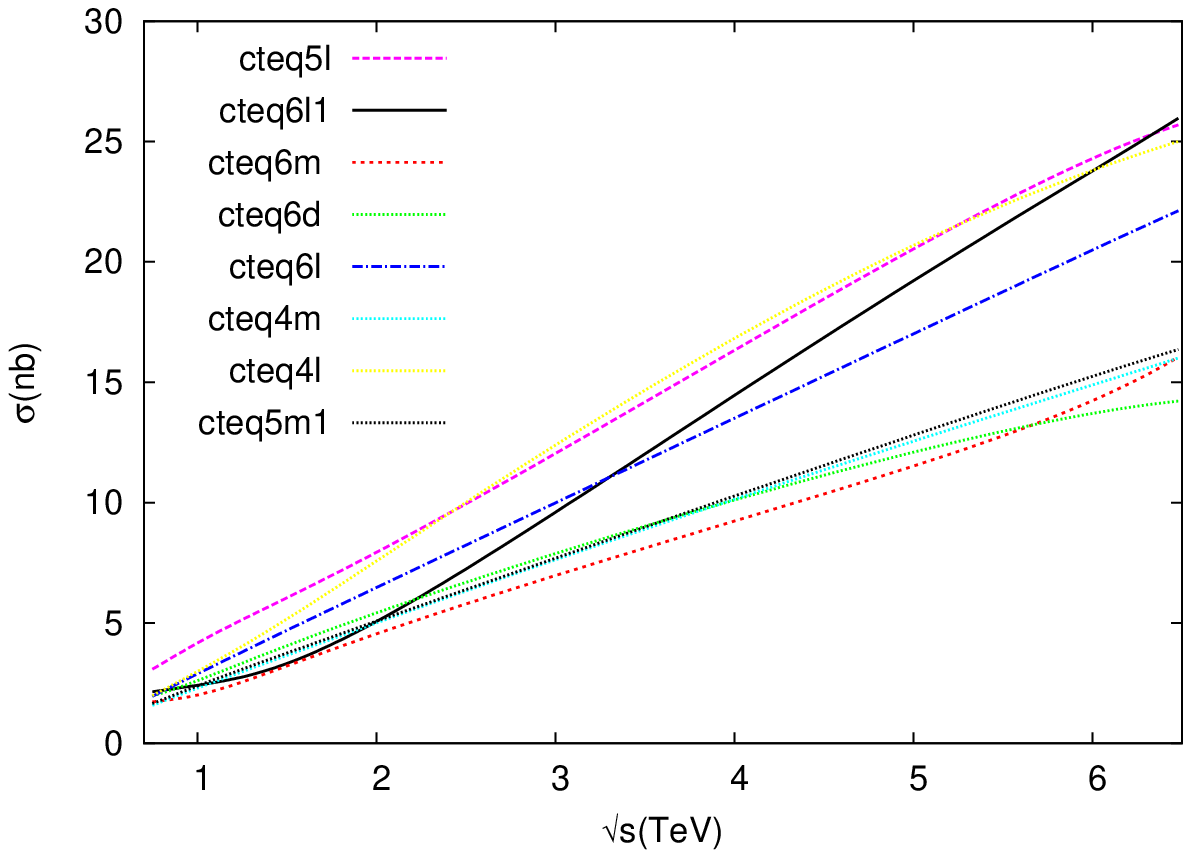}\caption{Total cross section versus center of mass energy values $(\sqrt{s})$
for different PDF distribution}

\end{figure}

\begin{figure}
\includegraphics[scale=0.6]{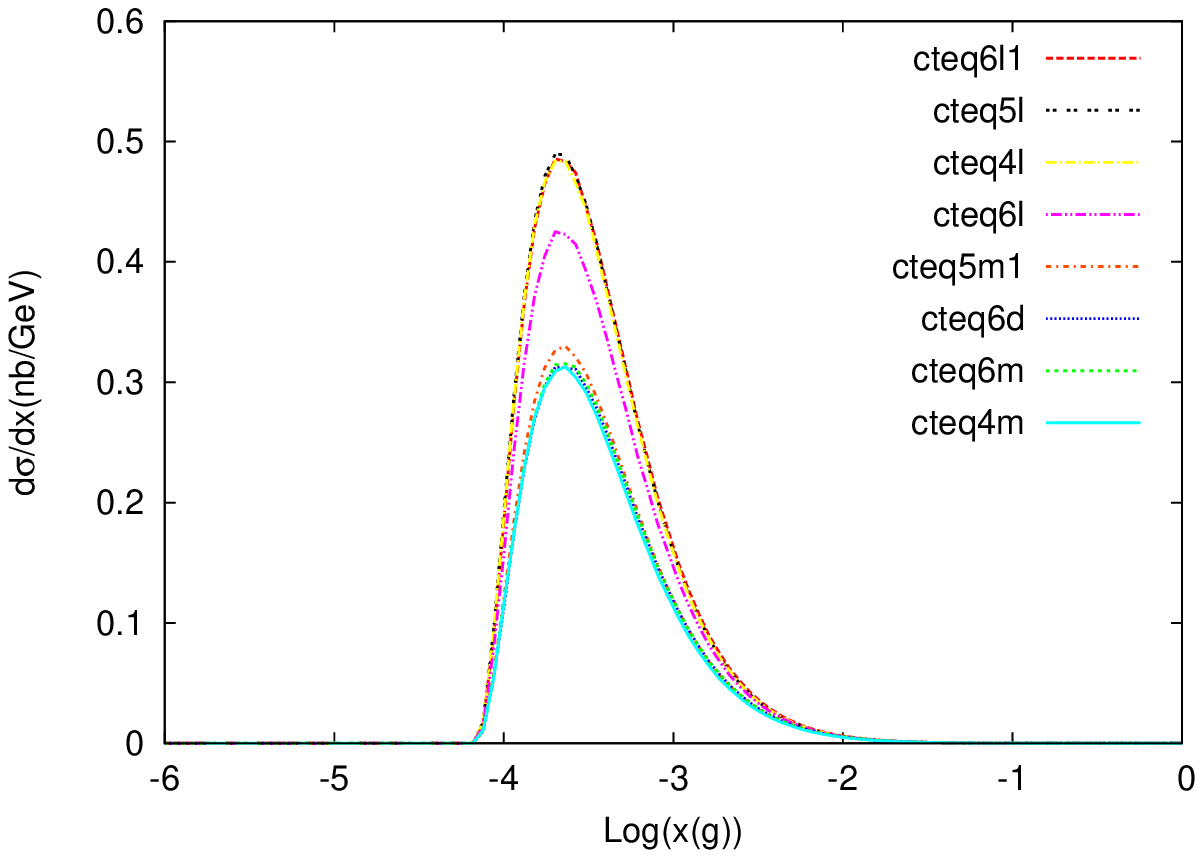}\includegraphics[scale=0.6]{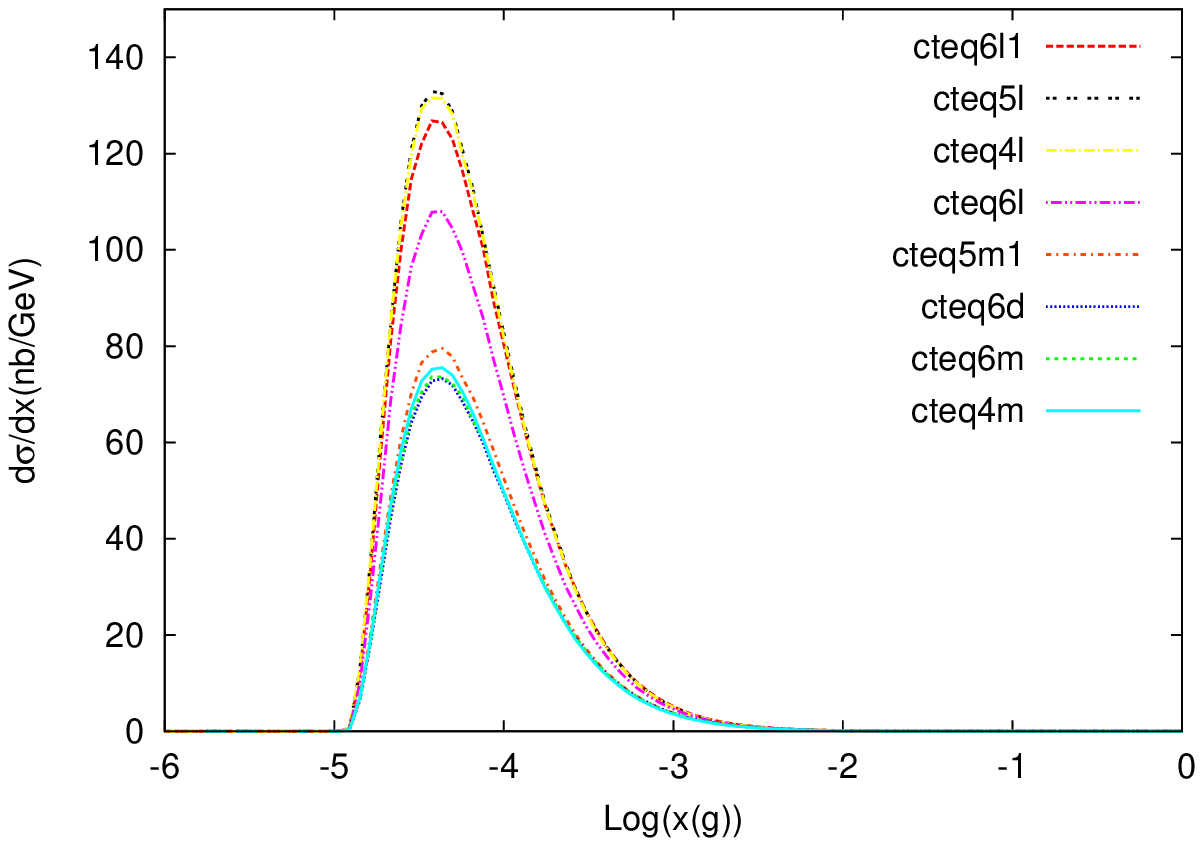}\caption{Differential cross section versus the $x(g)$ in $b\bar{b}$ (left)
and $c\bar{c}$ (right) for LHeC Type-1}

\end{figure}

\begin{figure}
\includegraphics[scale=0.6]{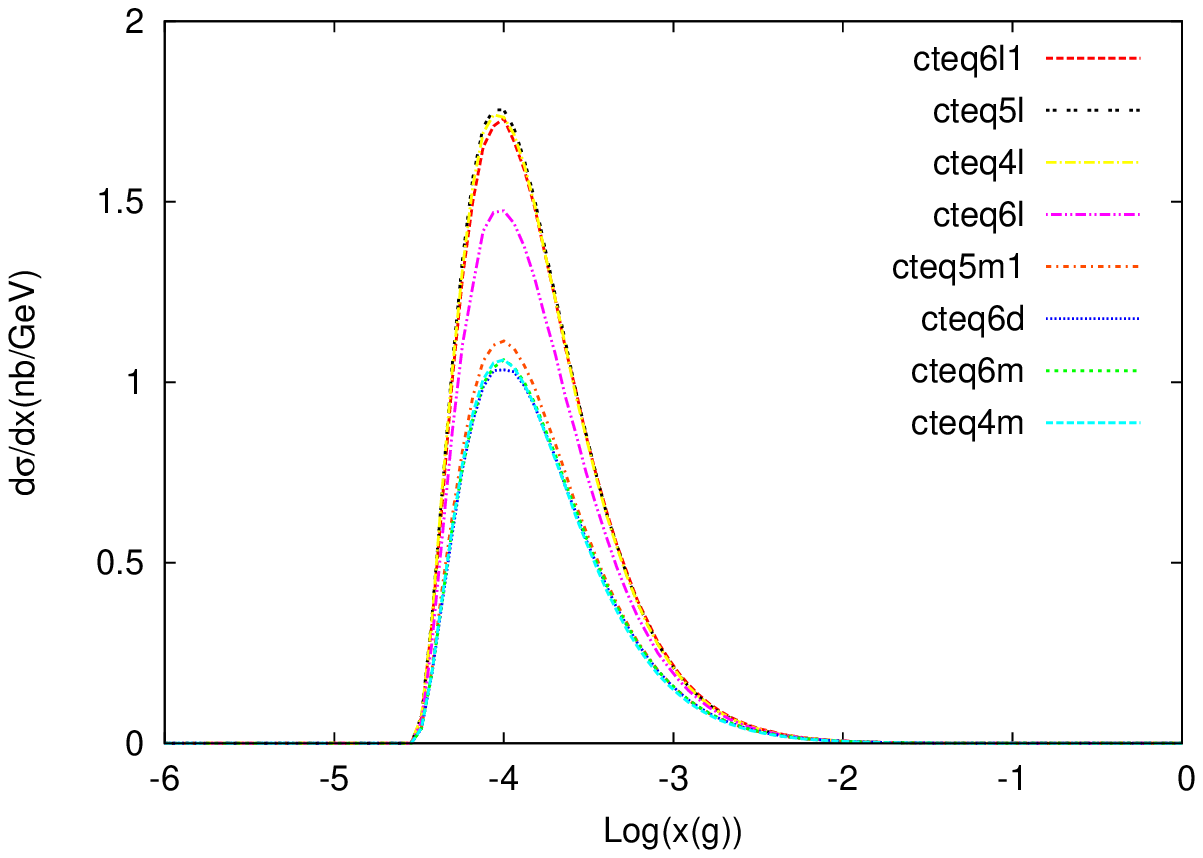}\includegraphics[scale=0.6]{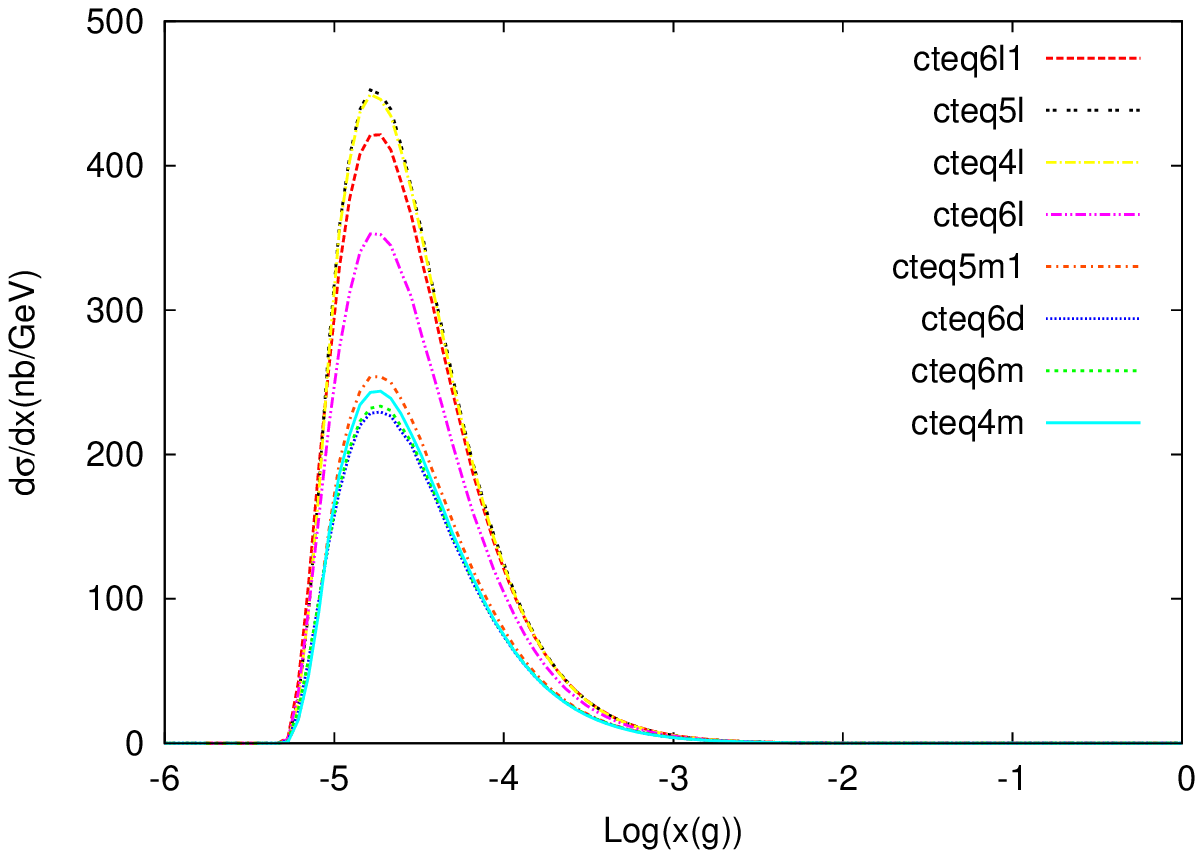}

\caption{Differential cross section versus the $x(g)$ in $b\bar{b}$ (left)
and $c\bar{c}$ (right) for LHeC Type-2}

\end{figure}

\begin{figure}
\includegraphics[scale=0.6]{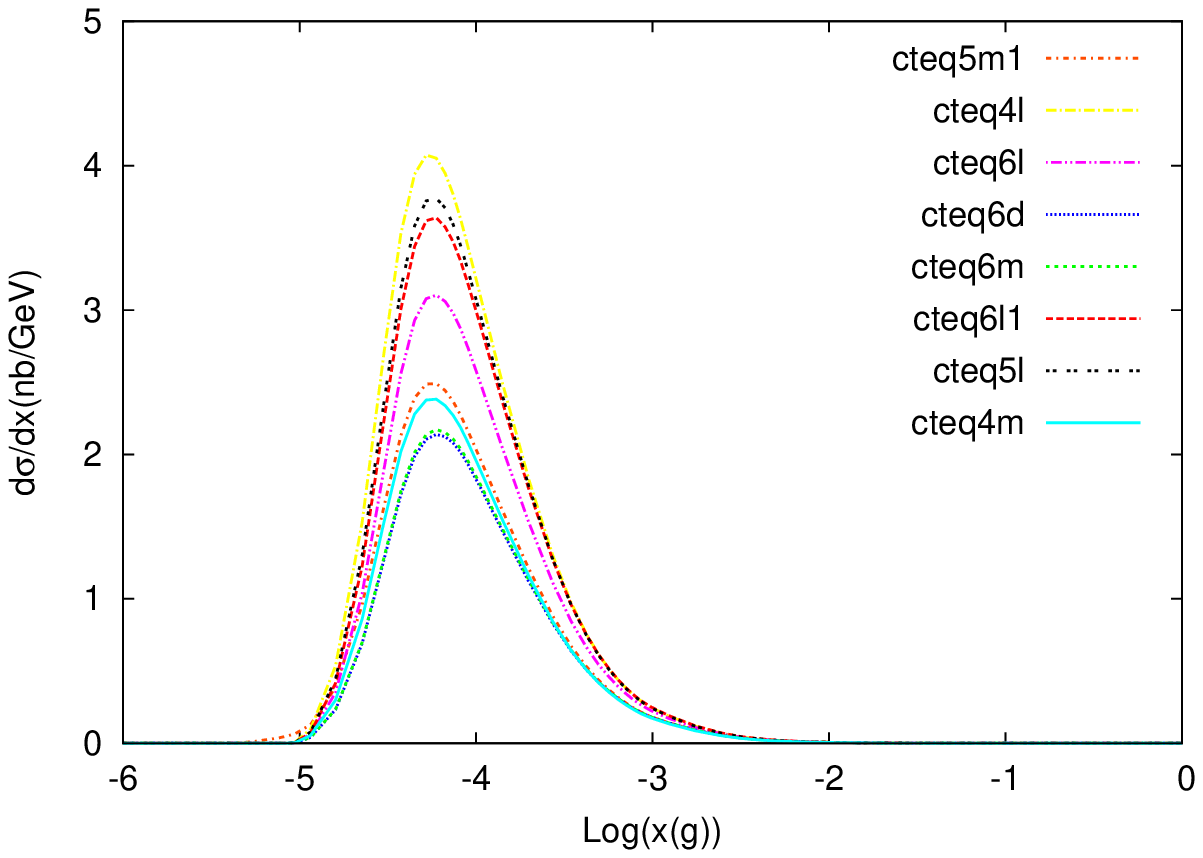}\includegraphics[scale=0.6]{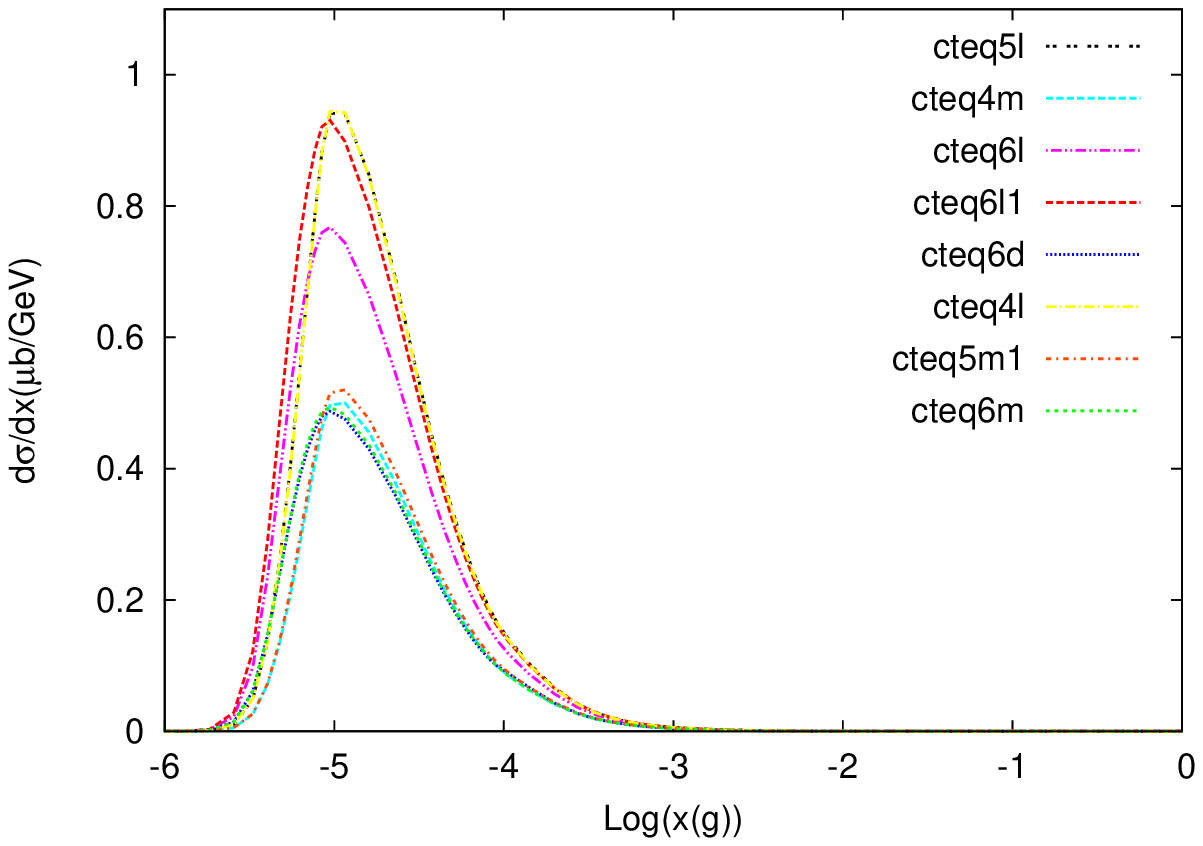}

\caption{Differential cross section versus the $x(g)$ in $b\bar{b}$ (left)
and $c\bar{c}$ (right) for $ep$ collider-1}

\end{figure}

\begin{figure}
\includegraphics[scale=0.6]{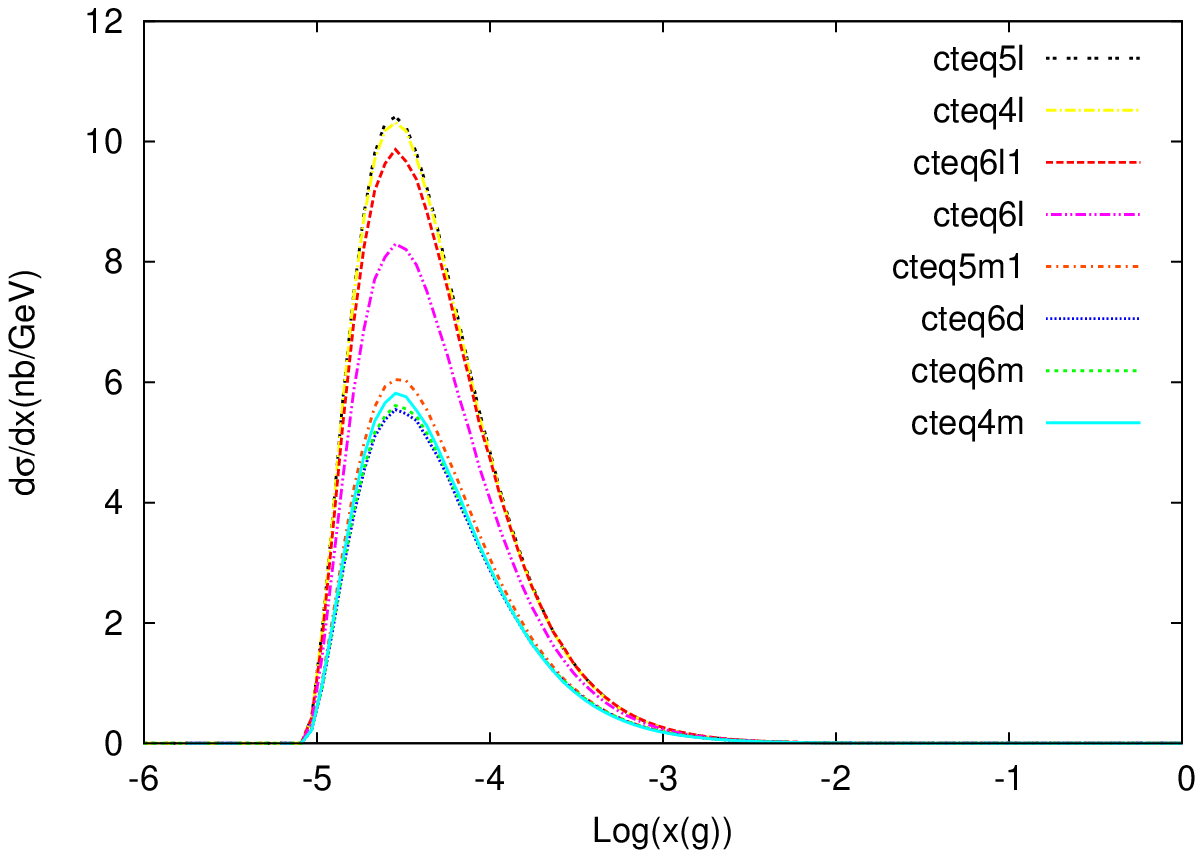}\includegraphics[scale=0.6]{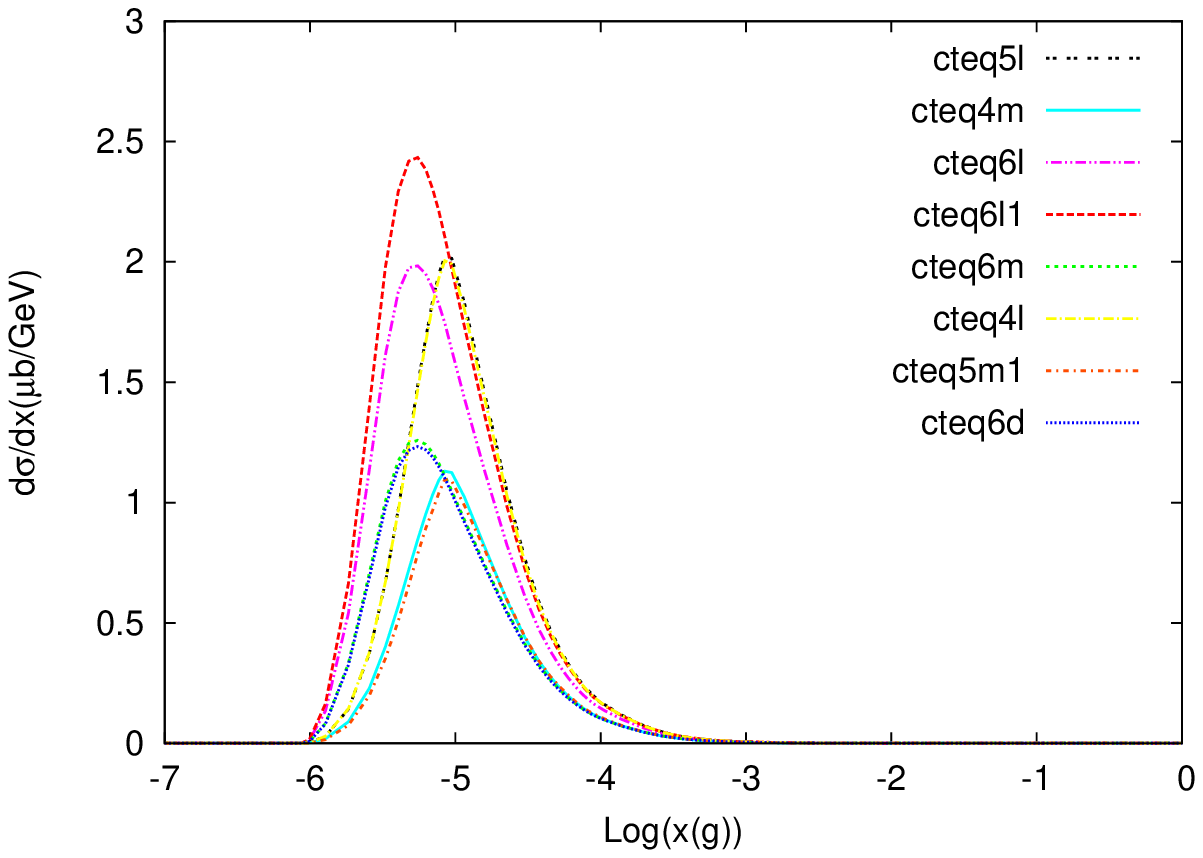}\caption{Differential cross section versus the $x(g)$ in $b\bar{b}$ (left)
and $c\bar{c}$ (right) for $ep$ collider-2}

\end{figure}

\end{document}